\documentclass[aps,prb,reprint,superscriptaddress,amsmath,amssymb,showpacs]{revtex4-1}%

\makeatletter
\newcommand*{\rom}[1]{\expandafter\@slowromancap \romannumeral #1@}
\makeatother

\renewcommand{\AA}{\mathcal{A}}
\newcommand{\CC}{\mathcal{C}}
\newcommand{\BB}{\mathcal{B}}
\newcommand{\DD}{\mathcal{D}}
\newcommand{\LL}{\mathcal{L}}
\newcommand{\dd}{\mathrm{d}}

\newcommand{\ii}{\text{i}}
\newcommand{\Tr}[1]{\text{Tr} \left\{ #1 \right\}}

\newcommand{\av}[1]{\left\langle #1 \right\rangle}

\newcommand{\expo}[1]{\text{exp}\left( #1 \right)}
\newcommand{\e}{\text{e}}
\newcommand{\tS}{\text{S}}

\newcommand{\tB}{\text{B}}

\newcommand{\pdagger}{{\phantom{\dagger}}}
\renewcommand{\bm}[1]{\textbf{\textit{#1}}}

\usepackage{braket}
\usepackage{import}
\usepackage{subfigure}
\usepackage{graphicx}
\usepackage{xcolor}
\usepackage{dsfont}
\usepackage{pstricks}
\usepackage{mathrsfs}
\usepackage{bbm}
\usepackage{nicefrac}

\begin{document}

\title{
Quantum thermodynamics of nonadiabatically driven systems: The effect of electron-phonon interaction}

\author{Jakob B\"atge}
\email{jakob.baetge@physik.uni-freiburg.de}
\affiliation{Institute of Physics, University of Freiburg, Hermann-Herder-Str. 3, D-79104 Freiburg, Germany}

\author{Amikam Levy} 
\email{amikam.levy@biu.ac.il}
\affiliation{Department of Chemistry, Bar-Ilan University, Ramat-Gan 52900, Israel }

\author{Wenjie Dou}
\email{douwenjie@westlake.edu.cn} 
\affiliation{Department of Chemistry, School of Science, Westlake University, Hangzhou, Zhejiang 310024, China }
\affiliation{Department of Physics, School of Science, Westlake University, Hangzhou, Zhejiang 310024, China }
\affiliation{Institute of Natural Sciences, Westlake Institute for Advanced Study, Hangzhou, Zhejiang 310024, China}

\author{Michael Thoss}
\email{michael.thoss@physik.uni-freiburg.de}
\affiliation{Institute of Physics, University of Freiburg, Hermann-Herder-Str. 3, 79104 Freiburg, Germany}
\affiliation{EUCOR Centre for Quantum Science and Quantum Computing,
University of Freiburg, Hermann-Herder-Str. 3, D-79104 Freiburg, Germany}

\begin{abstract}
 In this work we study the effects of nonadiabatic external driving on the thermodynamics of an electronic system coupled to two electronic leads and to a phonon mode, with and without damping.   In the limit of slow driving, we establish nonadiabatic corrections to quantum thermodynamic quantities. In particular, we study the first-order correction to the electronic population, charge-current, and vibrational excitation using a perturbative expansion, and compare the results to the numerically exact hierarchical equations of motion (HEOM) approach. Furthermore, the HEOM analysis spans both the weak and strong system-bath coupling regime and the slow and fast driving limits. We show that the electronic friction and the nonadiabatic corrections to the charge-current provide a clear indicator for the Franck-Condon effect and for non-resonant tunneling processes. We also discuss the validity of the approximate quantum master equation approach and the benefits of using HEOM to study quantum thermodynamics out of equilibrium.

\end{abstract}
 
\maketitle

\section{Introduction}

The rapidly growing field of single-molecule/atomic electronics \cite{sun2014single,metzger2015unimolecular,thoss2018perspective,gehring2019single,xin2019concepts} has implications for the development of thermoelectric devices \cite{reddy2007thermoelectricity,nozaki2014quantum,benenti2017fundamental}, molecular diodes \cite{elbing2005single,capozzi2015single}, optoelectronic molecular switches \cite{van2010charge,zhang2015towards,jia2016covalently}, molecular sensors \cite{wang2014frontispiece,zhao2014single}, computing devices and more. It further enables the study of thermodynamic properties in the quantum regime \cite{kosloff2013quantum,binder2018thermodynamics}, including entropy production, and energy and charge transport on the atomic scale. Miniaturizing electronic devices reveals the significance of physical phenomena that are absent in the thermodynamic limit and bulk materials such as the Coulomb and the Franck-Condon blockade, quantum interference, and quantum correlations. 

Theoretical models exploring these effects and treating strong system-lead couplings with stationary Hamiltonians have been studied extensively in the literature using different approaches including numerically exact methods such as the multilayer multiconfiguration time-dependent Hartree approach
\cite{wang2009numerically,wang2013multilayer,wang2018multilayer}, path integral and quantum Monte Carlo methods \cite{prldata,PhysRevB.96.155126,segal2010numerically,werner2009diagrammatic}, the hierarchical equations of motion (HEOM) \cite{tanimura2006stochastic,Schinabeck2016,erpenbeck2019hierarchical,Jin2008,Haertle2013} or approximate approaches such as the numerical renormalization group \cite{cohen2011memory,wilner2015sub,kidon2018memory,nrgreview,anders2008steady,heidrich2009real}, nonequilibrium Green’s function \cite{inelastic,mishaPRBfriction,PhysRevLett.114.080602,erpenbeck2015effect,hartle2008multimode}, scattering theory \cite{vonOppenPRB,bruch2018landauer,beilstein}, and mapping techniques \cite{levy2019complete,strasberg2016nonequilibrium,PhysRevE.95.032139,PhysRevB.97.205405,gelbwaser2015strongly,katz2016quantum}.  Yet the effects of nonadiabatic driving on the transport properties have received much less attention, mainly because of the difficulties of solving such complex dynamics. And this is despite the fact that the ultimate goal of the field is to control and manipulate microelectronic devices to perform a certain task. The control itself is typically achieved by applying external fields to the system that can be expressed theoretically using time dependent Hamiltonians
\cite{Bruch2016,Karimi2016,Solfanelli2020,haughian2018quantum,ochoa2018quantum,Kuperman2020,Honeycurch2019,Preston2020,Restrepo2019,Liu2021}
.

One of the main characteristics of nonadiabatic driving of an open quantum system is electronic  \cite{beilstein,PhysRevB.96.104305,PhysRevB.97.064303,PhysRevLett.119.046001} and quantum \cite{kosloff2002discrete,plastina2014irreversible,levy2018quantum} friction. These phenomena are directly related to the dissipation into the environment of the excitations and coherence induced by the external driving. The environment itself can be composed of several leads with different temperatures and/or chemical potentials that impose nonequilibrium dynamics on the system, which manifest as energy, charge and entropy flows in and out of the system. In Ref.~\onlinecite{Dou20} we established a thermodynamic description capturing the effects of finite time driving on quantum impurity models out of equilibrium. We employed a perturbative expansion around the adiabatic, slow-driving, limit and explored the corrections to thermodynamic properties such as the entropy production and energy flows. The analysis was applied to the driven Anderson impurity model that exhibits Coulomb-blockade signatures in nonadiabatic correction of thermodynamic quantities. 

In this work, we present a comprehensive study exploring the effects of electron-phonons/photons couplings on the dynamics and thermodynamics of a driven resonant-level-model. To unravel the phonons’ contribution, we first study the driven resonant-level-model coupled to two electric leads in and out of equilibrium. We then add, layer by layer, the contribution of a single phonon and finally assume that this phonon is further coupled to a phononic thermal bath. The analysis is based both on the perturbation expansion introduced in Ref.~\onlinecite{Dou20} combined with a quantum master equation (QME) approach \cite{bookNitzan}, and on the numerically exact HEOM approach. The latter allows to expand the study to the strong system-environment coupling regime and to fast driving – much faster than the perturbation theory is valid for. We note in passing that an approximate QME can be applied in certain cases to fast driving as well. It requires deriving a time dependent master equation \cite{dann2018time} or using perturbative expansion of the dissipative part of the QME \cite{levy2021response}.

The comparison between the results obtained from the QME and the HEOM reveals the role of co-tunneling processes in electronic friction with and without the presence of electron-phonon couplings. This comparison also helps us to understand to what extend the approximate QME approach is reliable when the open system is externally driven. We further show that nonadiabatic correction to thermodynamic quantities provides signatures to non-resonant processes in the resonant-level-model and to the Franck-Condone principle when phonons are included in the model.

The paper is organized as follows. In Section \ref{sec:general_qtd}, we briefly review quantum thermodynamic properties of open quantum system in the presence of external driving fields. In Sec.~\ref{sec:qtd_el_ph_coupled}, we introduce our model with electron-phonon couplings and the HEOM approach as well as the QME technique. In Sec.~\ref{sec:num_results}, we analyze the results with focus on the nonadiabatic limit. Finally, we conclude in Sec.~\ref{sec:conclusion}.

\section{quantum thermodynamic properties}   
\label{sec:general_qtd}

Here, we briefly review quantum thermodynamics for a general out of equilibrium case. The total system consists of a dot, left and right baths, and the interactions between them,
\begin{eqnarray}
H = H_s + \sum_{\alpha=L, R} (H_\alpha + H_{I\alpha}).
\end{eqnarray}
Without driving, we assume that the total system reaches a steady state
\begin{eqnarray}
\rho_{ss} =  e^{-\bar \beta (H- Y)} / \Omega,
\end{eqnarray}
where $\Omega = tr (e^{-\bar \beta ( H- Y)} ) $ is the normalization factor with the reduced inverse temperature $\bar \beta$ and $Y$ is the particle (or heat) transport operator.\cite{Ness2017} To include the driving let us consider that the Hamiltonian depends on a set of parameters $R_i$, which vary slowly as a function of time. With the steady state density operator, we can define the thermodynamic properties, such as work rate and current in the adiabatic limit:\cite{Dou20}
\begin{eqnarray}
\dot W^{(1)} &=& \sum_i \dot R_i tr (\partial_i H  \rho_{ss} ), \\
I_{\alpha}^{(0)} &=& tr ( - i [ H, N_{\alpha}] \rho_{ss}),  
\end{eqnarray}
where we denote $\frac{\partial}{\partial R_i} \equiv \partial_i$, $\dot R_i$ is the driving speed, and $N_{\alpha}$ is the number operator for the $\alpha=L,R$ leads. We have used $^{(n)}$ to denote nth order in driving speed. In case only the dot Hamiltonian depends on external parameter and the dot only consists of one level ($H_s = \epsilon_d d^\dagger d$), we have
\begin{eqnarray}
\dot W^{(1)} = \dot \epsilon_d N^{(0)}, \\
N^{(0)} = tr (d^\dagger d  \rho_{ss} ).
\end{eqnarray}
Here $N^{(0)}$ is the dot population in the adiabatic limit and $\dot \epsilon_d$ is the driving speed of the dot level energy (see also Sec. \ref{sec:qtd_el_ph_coupled}).

Now we consider the case where we have finite driving speeds. The equation of motion for the density reads
\begin{eqnarray}
\partial_t \rho   +  \sum_i \dot{ R}_i  \partial_i \rho  = - i[ H, \rho ].  
\end{eqnarray}
We further expand the density into a series in the power of driving speed, 
\begin{eqnarray}
\rho = \rho^{(0)} + \rho^{(1)} + \rho^{(2)} + \cdots .
\end{eqnarray}
With the steady state $\rho^{(0)} = \rho_{ss}$, we can solve for the first order nonadiabatic correction to the density:
\begin{eqnarray}
\rho^{(1)} = \int_0^{\infty} e^{-iHt}  \sum_j \dot{ R}_j  \partial_j \rho_{ss} e^{iHt} dt  .
\end{eqnarray}
Here, we have used the Markovian approximation that is consistent with the adiabatic limit, i.e. the driving is much slower than the relaxation of the system.\cite{Dou18,Dou20} Using the first order correction to the state, we can calculate the nonadiabatic correction to the thermodynamic quantities: 
\begin{eqnarray}
\dot  W^{(2)} = \sum_{ij} \dot R_i \dot R_j  \int_0^{\infty} tr ( e^{-iHt}  \partial_j \rho_{ss} e^{iHt} \partial_i H )dt, \\
 I_{\alpha}^{(1)} = tr ( - i [ H, N_{\alpha}] \rho^{(1)}).
\end{eqnarray}
The nonadiabatic correction $\dot  W^{(2)}$ is the dissipative work, and we can further introduce a frictional tensor $\gamma_{ij}$, such that $\dot  W^{(2)} =\sum_{ij} \dot R_i \gamma_{ij} \dot R_j $. 

Note that the analysis above is general. We have applied such analysis to the non-interacting electronic systems to obtain analytical results.\cite{Dou18,Dou20} For the interacting systems, analytical results are not available. Below we will apply the HEOM approach to analyze quantum thermodynamic properties when including electron-phonon interactions.

\section{Model and Methods}
\label{sec:qtd_el_ph_coupled}

\begin{figure}
    \centering
    \includegraphics{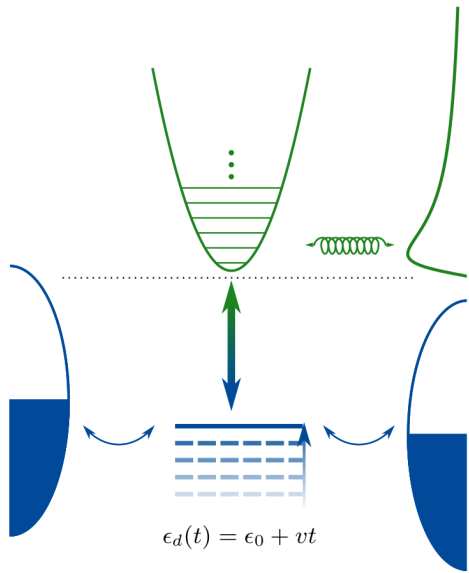}
    \caption{Sketch of the model consisting of one driven electronic state interacting with a single vibrational mode coupled to two electronic leads and
one vibrational heat bath.}
    \label{fig:model}
\end{figure}

For our scenario of vibrationally coupled and externally driven electron transport through a nanostructure (see Fig.~\ref{fig:model}), the Hamiltonian is given by (using units where $\hbar=1$) 
\begin{subequations}
\label{eq:general_Hamiltonian}
\begin{align}
H =& \epsilon_d (t) d^\dagger d + \lambda  (a^\dagger + a) d^\dagger d + \Omega a^\dagger a   \label{eq:Hamiltonian_S} \\
 & +\sum\limits_{k\alpha} \nu^\pdagger_{k\alpha} ( c^\dagger_{k\alpha} d + d^\dagger  c_{k\alpha}^\pdagger  )  + \sum\limits_{k\alpha} \epsilon^\pdagger_{k\alpha}  c^\dagger_{k\alpha} c^\pdagger_{k\alpha}  \label{eq:Hamiltonian_SE} \\
 & + \sum\limits_{j} \xi^\pdagger_{j} \left(  {b}_{j}^\dagger + {b}_{j}^\pdagger \right)  \left(a^\dagger +a\right)  + \sum\limits_{j} \omega^\pdagger_{j} {b}_{j}^\dagger {b}_{j}^\pdagger. \label{eq:Hamiltonian_E}
\end{align}
\end{subequations}
Here, the system part of the Hamiltonian consists of an electronic state with an externally controlled energy $\epsilon_d(t)$, a harmonic mode with frequency $\Omega$ and an adiabatic coupling of the electronic state to the harmonic mode with a coupling strength  $\lambda$.
The electronic state and the harmonic mode are addressed by their creation (annihilation) operator $d^\dagger (d)$ and $a^\dagger (a)$.  We further assume that we have linear driving for the electronic state energy
$\epsilon_d (t)  = \epsilon_0 + v t,$
where $v$ denotes the driving velocity.
On the one hand, the environment includes two (left and right) macroscopic electron reservoirs, where the $k$-th electronic state in the electron reservoir $\alpha\in\{L,R\}$ with an energy $\epsilon_{k \alpha}$ is addressed by its creation (annihilation) operator $c_{k\alpha}^\dagger (c_{k\alpha}^\pdagger)$ and the corresponding coupling to the system is specified by $\nu_{k\alpha}$. Via the chemical potentials of the electron reservoirs $\mu_\alpha$, we can apply bias voltages $\Phi=\mu_L-\mu_R$ to the system.
On the other hand, the environment also includes a microscopic heat bath, where the $j$-th harmonic mode with a frequency $\omega_j$ has the creation (annihilation) operator $b_j^\dagger$  $(b_j^\pdagger)$ and the coupling strengths $\xi_j$.
The entire environment is considered to have a constant temperature $T$.

The influence of the environments onto the system is further characterized by their respective spectral densities   
 \begin{subequations}
  \begin{align}
        \Gamma_{\alpha}(\epsilon) =& 2 \pi \sum\limits_{k} |\nu^\pdagger_{k \alpha}|^2 \delta(\epsilon-\epsilon^\pdagger_{k \alpha}) = \Gamma_{\alpha} \frac{D_{\alpha}^2}{D_{\alpha}^2+(\epsilon-\mu_{\alpha})^2}  ,
    \label{eq:fermionic_spectral_density_general_definition}
\\
    \Lambda(\omega) =& \pi \sum\limits_{j} |\xi^\pdagger_{j}|^2 \delta(\omega-\omega^\pdagger_{j }) = \Lambda \frac{\omega}{\Omega} \frac{ \omega^2_{c}}{\omega_{c}^2+\omega^2}.
    \label{eq:bosonic_spectral_density_general_definition}
  \end{align}
 \end{subequations}  
Here,  $\Gamma_\alpha$ denotes the coupling strengths of the electron reservoirs and $D_\alpha$ the bandwidth of the Lorentzian shaped spectral density. In the following, we effectively use the wide band approximation by the choice of $D=30\,$eV and assume symmetrically coupled reservoirs with $\Gamma_L=\Gamma_R=\frac{\Gamma}{2}$. In addition, the spectral density of the heat bath is Ohmic with a Lorentzian cut-off with the cut-off frequency $\omega_c$ and a coupling strength $\Lambda$. Throughout this work, we choose $\omega_c=\Omega$.

The isolated system can be diagonalized using the so-called small Polaron transformation \cite{Lang1963,Mahan2013}, leading to the renormalized electronic state energy of $\overline{\epsilon}_d(t)=\epsilon_d(t) - \frac{\lambda^2}{\Omega}$.

Below, we will apply numerical exact solution from HEOM to study the thermodynamics for such a Hamiltonian.

\subsection{HEOM with electron-phonon couplings as well as time dependent driving}

In the following, we present the most important steps of the derivation of the  
numerically exact HEOM approach for the model under investigation. Thereby, we closely follow Refs.\ \onlinecite{Schinabeck2016} and \onlinecite{Baetge2021}. More detailed derivations 
are presented in Refs. \onlinecite{Tanimura1989,Jin2008}.

The derivation of the HEOM is based on the system-environment partitioning (see Eq.~\eqref{eq:general_Hamiltonian}). The central quantity of the approach is the reduced density matrix $\rho(t)$ of the system, where the bath degrees of freedom are traced out. The influence of the environment on the system dynamics is taken into account by the Feynman-Vernon influence functional.
For our model Hamiltonian, all information about system-environment coupling is encoded in the two-time correlation functions of the free environments 
 \begin{subequations}  
 \begin{align}
    \tilde{C}(t-\tau)=&\sum\limits_{j} |\xi^\pdagger_{j}|^2 \av{   {b}_{j}^\dagger (t) {b}_{j }^\pdagger(\tau)  +  {b}_{j }^\pdagger(t) {b}_{j}^\dagger(\tau) } ,
    \\
    C^s_{\alpha}(t-\tau)=&\sum\limits_{k } |\nu_{k \alpha}|^2 \av{ 
 c^{s}_{k \alpha}(t)  c^{\bar{s}}_{k \alpha}(\tau) } ,
 \end{align}
 \end{subequations}  
 which are determined by the respective spectral densities
\begin{align}
  \tilde{C}(t) = & \int_0^\infty d\omega\frac{\Lambda(\omega)}{\pi} \left[\text{coth}\left(\frac{\beta \omega}{2}\right)\text{cos}(\omega t) - i \text{ sin}(\omega t)\right],
      \\
  C^s_{\alpha} (t)=&\frac{1}{2 \pi} \int_{-\infty}^\infty \dd \epsilon\, \e^{s \ii \epsilon t/\hbar} \Gamma_{\alpha} (\epsilon) f [s (\epsilon-\mu_\alpha)].
\label{eq:C_FT}
\end{align}
Here, $f(\epsilon)=\left( \expo{\beta \epsilon} +1 \right)^{-1}$ denotes the Fermi distribution and $\beta=\frac{1}{T}$ the inverse temperature.
Furthermore, the notations $c^+=c^\dagger$, $c^-=c$ and $\bar{s}=-s$ are employed.
To derive a closed set of equations of motion within the HEOM method, all correlation functions are expressed by sums over exponentials.\cite{Jin2008} To this end, the Fermi as well as the Bose distribution  are represented by sum-over-poles schemes employing Pad\'e decompositions.\cite{Ozaki2007,Hu2010,Hu2011}
Thus, the correlation functions of the free baths are given by $\tilde{C}(t)=\Lambda \sum_{p=0}^{p_\text{max}} \tilde{\eta}_p \e^{-\tilde{\gamma}_p t} $ respectively $C^s_{\alpha}(t)=\Gamma_\alpha \sum_{q=0}^{q_\text{max}} \eta_{\alpha,q}  \e^{-\gamma_{\alpha,s,q} t}$. 
Therefore one obtains the HEOM in the form of
\begin{alignat}{2}
   \frac{\partial}{\partial t} \rho^{(m|n)}_{\bm{g}|\bm{h}} =& -\left( i \LL_\tS + \sum_{l=1}^m \tilde{\gamma}_{g_l} + \sum_{l=1}^n \gamma_{h_l} \right) \rho^{(m|n)}_{\bm{g}|\bm{h}}
   \nonumber \\
   &- \sum_{h_x} \AA_{h_x} \rho^{(m|n+1)}_{\bm{g}|\bm{h}^+_x} -\sum_{l=1}^n (-1)^l \CC_{ h_l}  \rho^{(m|n-1)}_{\bm{g}|\bm{h}^-_l}
   \nonumber \\
   &+ \sum_{g_x} \BB_{g_x} \rho^{(m+1|n)}_{\bm{g}^+_x|\bm{h}} +\sum_{l=1}^m \DD_{ g_l}  \rho^{(m-1|n)}_{\bm{g}^-_l|\bm{h}},
   \label{eq:general_HEOM}
\end{alignat}
with the multi-indices $g=(p)$ and $h=(\alpha,s,q)$, the notation for the multi-index vectors ${\bm{v}} = {v_1 {\cdot}{\cdot}{\cdot}v_p}$, ${\bm{v}^+_x}= {v_1 {\cdot}{\cdot}{\cdot}v_p v_x}$, and ${\bm{v}^-_l} = {v_1 {\cdot}{\cdot}{\cdot}v_{l-1}v_{l+1}{\cdot}{\cdot}{\cdot}v_p}$,
and $\LL_\tS O = [H_\tS,O]$.
The superoperators $\AA_{h}$, $\CC_{h}$, $\BB_{g}$ and $\DD_{g}$ read
 \begin{subequations}  
\begin{align}
 \AA_{h} \rho^{(m|n)}_{\bm{g}|\bm{h}}=&  \Gamma_{\alpha_h} \left( d^{s_{h}}  \rho^{(m|n)}_{\bm{g}|\bm{h}} + (-1)^{(n)} \rho^{(m|n)}_{\bm{g}|\bm{h}} d^{s_{h}} \right),\\
 \BB_{g} \rho^{(m|n)}_{\bm{g}|\bm{h}}=&  \Lambda  \left[ \left(a^\dagger +a\right) , \rho^{(m|n)}_{\bm{g}|\bm{h}}\right],\\
 \CC_{h} \rho^{(m|n)}_{\bm{g}|\bm{h}}=&\, (-1)^{n} \eta_{h}   {d^{\bar{s}_{h}} }  \rho^{(m|n)}_{\bm{g}|\bm{h}} - \eta^*_{ \bar{h}}  \rho^{(m|n)}_{\bm{g}|\bm{h}} {d}^{\bar{s}_{h}}, \\
 \DD_{g} \rho^{(m|n)}_{\bm{g}|\bm{h}}=&\, \tilde{\eta}_{g}   \left(a^\dagger +a\right)  \rho^{(m|n)}_{\bm{g}|\bm{h}} - \tilde{\eta}^*_{ g}  \rho^{(m|n)}_{\bm{g}|\bm{h}} \left(a^\dagger +a\right).
 \label{eq:general_HEOM_upbuilding_operators}
\end{align}
 \end{subequations}  
Due to system-environment interaction, these superoperators couple the different levels of the hierachy.

Here, $\rho^{(0)} \equiv \rho$ represents the reduced density matrix and $\rho^{(m|n)}_{\bm{g}|\bm{h}}$ $(n+m>0)$ denote auxiliary density matrices, which describe environment-related observables such as, e.g., the charge-current
\begin{alignat}{2}
 I_{\alpha} &= - e \av{\frac{d N_\alpha }{d t}} &&=  e\,\Gamma_{\alpha} \sum_{h_\alpha} s_h \Tr{ d^{\bar{s}_{h_\alpha}} \rho^{(0|1)}_{\pdagger|{h_\alpha}} }.
\end{alignat}

The importance of the auxiliary density operators to the system dynamics is estimated by assigning them the importance values,\cite{Haertle2013,Baetge2021} 
 \begin{align}
	 \mathcal{I} \left( \rho^{(m|n)}_{\bm{g}|\bm{h}}\right) =&  \left|\prod\limits_{l=1}^{n }\frac{\Gamma}{\sum\limits_{a\in\{1..{l}\}}\hspace{-.2cm} \text{Re}\left[\gamma_{h_{a}}\right]}   \frac{\eta_{h_{l}}}{\text{Re}\left[\gamma_{h_{l}}\right]} \right|\nonumber \\
	 &\times \left| \prod\limits_{l=1}^{m }\frac{\Lambda}{\sum\limits_{a\in\{1..{l}\}}\hspace{-.2cm} \text{Re}\left[\gamma_{g_{a}}\right]}  \frac{\eta_{g_{l}}}{\text{Re}\left[\gamma_{g_{l}}\right]} \right|.
	 \label{eq:importance_estimate}
\end{align}
In the calculations presented in this paper, the results are quantitatively converged for truncation of the hierarchy at level $m=2$ and $n=2$, neglecting auxiliary density operators having an importance value smaller $10^{-9}$.

\subsection{Nonadiabatic corrections   from HEOM calculations}

In order to calculate nonadiabatic corrections   to physical properties of the system, such as the populations and the currents, using HEOM calculations, and to compare it to linear response based on QME calculations we performer two types of calculations. 
First, we prepare the system in a stationary state, where the electronic state energy is much lower than both chemical potentials of the leads. Starting from this stationary state, we increase the energy with a constant velocity and track the time-evolution of the system $\rho(t)$. Thereby, we ensure that the time-evolution in the energy range of interest is independent of the initial electronic state energy. 
Second, we calculate the system steady states $\rho_{ss}(\epsilon_d)$ for the electronic state energies in our range of interest.

 Using these two calculations we quantify the nonadiabatic correction of a system observable $\av{O}$ from the HEOM calculations by 
\begin{equation}
\delta \av{O} = \frac{ \av{O}(t) - \av{O}_{ss}} {v } .
\label{eq:def_nonadiabatic_correction}
\end{equation}
Here,  $\av{O}(t)=\Tr{O \rho(t)}$ is the expectation value of the driven system, and the steady state expectation value $\av{O}_{ss}~=~\Tr{O \rho_{ss}(\epsilon_d(t))}$ is evaluated at the instantaneous energy $\epsilon_d(t)$ of the driven system. Thereby, the nonadiabatic correction $\delta\av{O}$ from the HEOM calculations includes corrections of first and higher order in the driving velocity.

In this study we will focus on the electronic friction represented by the nonadiabatic correction of the electronic population $\av{d^\dagger d}$ 
\begin{eqnarray} \label{eq:gammadd}
\gamma = \delta \av{d^\dagger d} = \frac{ \av{d^\dagger d}(t) - \av{d^\dagger d}_{ss}} {v } 
\end{eqnarray}
as well as the nonadiabatic correction to the current 
\begin{eqnarray}
\delta I = \frac{ I(t) - I_{ss}} {v},
\end{eqnarray}
and  the corrections to the vibrational excitation $\delta \av{a^\dagger a}$ that can be calculated in a similar manner. 

\subsection{Quantum master equation analysis}

In the limit of weak system-leads couplings, a reduced description of the system can be achieved by solving the quantum master equation:
\begin{eqnarray}
\partial_t \rho_s = - \dot \epsilon_d \partial_{\epsilon_d} \rho_s   - i [ H_s , \rho_s]  - \mathcal{D} \rho_s.
\end{eqnarray}
Here, $\mathcal{D}=\mathcal{D}_L + \mathcal{D}_R $ is a super-operator representing the (left and right) system-lead couplings and is responsible for dissipation and decoherence processes.  In the weak coupling limit, $\mathcal{D}$ can be expressed in Lindblad form \cite{lindblad1976generators}. Below, we denote $\mathcal{L}(\cdot) = i [ H_s , \cdot] + \mathcal{D}(\cdot) $. Similar to the analysis in Sec.~\ref{sec:general_qtd}, we can expand the system density matrix into the power of the driving speed. When matching the order from both sides of the above equation, we obtain
\begin{eqnarray}
\partial_t \rho_s^{(0)} =  - \mathcal{L} \rho_s^{(0)}, \\
\partial_t \rho_s^{(1)} =  -\dot \epsilon_d \partial_{\epsilon_d} \rho_s^{(0)} - \mathcal{L} \rho_s^{(1)}, 
\end{eqnarray}
where $\rho_s^{(0)}$ is the steady state solution of the system density in the adiabatic limit, which can be obtained by solving for the  non trivial solution of $\mathcal{L} \rho_s^{(0)} =0$. With $\rho_s^{(0)}$ at hand, we can proceed to solve for the nonadiabatic corrections,  
\begin{eqnarray}
\rho_s^{(1)} =  \dot \epsilon_d \int_0^{t}  e^{-\mathcal{L}t'}  \partial_{\epsilon_d} \rho_s^{(0)}  dt'. 
\end{eqnarray}
In the limit where the driving is slower than the timescale for the system response, we can invoke the Markovian approximation for the first order correction to the density matrix: 
\begin{eqnarray}
\rho_s^{(1)} =  \dot \epsilon_d \int_0^{\infty}  e^{-\mathcal{L}t} \partial_{\epsilon_d} \rho_s^{(0)}  dt  = - \dot \epsilon_d  \mathcal{L}^{-1} \partial_{\epsilon_d} \rho_s^{(0)}.
\end{eqnarray}
In the above equation, since $\partial_{\epsilon_d} \rho_s^{(0)} $ is traceless, $\mathcal{L}^{-1}$ can be acted on $\partial_{\epsilon_d} \rho_s^{(0)}$ properly. 
In the weak coupling limit, the population and electron current can be express using the system density operator alone, 
\begin{eqnarray}
N  =  Tr (d^\dagger d \rho_s), \\
I_{\alpha} = tr( d^\dagger d \mathcal{D}_\alpha \rho_s ).  
\end{eqnarray}
Again, $\mathcal{D}_\alpha$ is a super-operator representing the couplings between the $\alpha = \{L, R\}$ lead and the system. 
Replacing $\rho_s$ by $\rho_s^{(0)}$ will give us the nonadiabatic correction to these quantities. The friction $\gamma$ is  related to the nonadiabatic correction to the population via
\begin{eqnarray}
N^{(1)}  =  \dot \epsilon_d  Tr (d^\dagger d \mathcal{L}^{-1} \partial_x \rho_s^{(0)}) = \dot \epsilon_d \gamma, 
\label{eq:QME_Friction}
\end{eqnarray}
and, similarly, the first order correction to the current is given by 
\begin{eqnarray}
I_\alpha^{(1)} =  tr( d^\dagger d \mathcal{D}_\alpha \rho_s^{(1)} ).
\end{eqnarray}
Below, we compare our weak coupling analytical results with the numerical results from the HEOM calculations and focus on the  friction, the vibrational excitation, and the charge current. 

\section{Results} 
\label{sec:num_results}

In the following, we discuss the response of electronic-vibrationally coupled systems under a linear drive of the electronic system energy. 
In Sec.~\ref{sec:NumRes:sub:Eq} and \ref{sec:NumRes:sub:Eq:elph}, we focus on the system with and without electronic-vibrational interaction at equilibrium  i.e. we do not apply any bias voltage. We continue to investigate the system response to the linear drive for situations in which a bias voltage is applied to the system in Sec.~\ref{sec:NumRes:sub:NonEq} and \ref{sec:Res:Bias_and_elph}. Next, in Sec.~\ref{sec:NumRes:sub:EnvDamp}, we study the effect of environmental damping of the vibrational mode on the system response. Finally, we discuss the response of the system for fast driving in which the linear response treatment is no longer valid, Sec.~\ref{sec:NumRes:sub:FastDrive}.

\subsection{Zero bias-voltage (equilibrium): resonant-level-model}
\label{sec:NumRes:sub:Eq}

\begin{figure}[!t]
 \centering
 \includegraphics{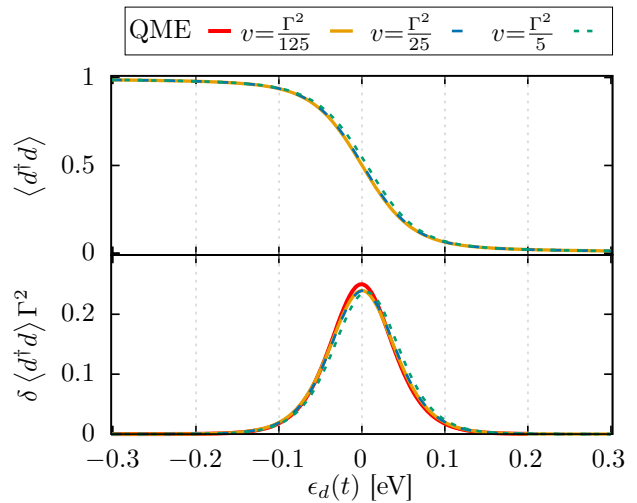}
 \caption{ Electronic population $\av{d^\dagger d}$ and the corresponding friction for a non-interacting system without bias voltage and for different driving velocities $v$ as a function of the time-dependent energy $\epsilon_d(t)=\epsilon_0+vt$. When the electronic state energy passes by the chemical potentials of the electron reservoirs, i.e. $\epsilon_d(t)=0$, we find a drop in the electronic population and a peak in the corresponding friction.
 Further parameters are  $k_\tB T=\Gamma=0.025$\,eV.}
 \label{fig:Equilibrium_non-interacting}
\end{figure}

We begin our investigation with the simple resonant-level-model. In this case, no bias-voltage is applied to the system and no  electronic-vibrational couplings are considered, i.e. $\Phi=0=\frac{\lambda}{\Omega}$.
As there is no current or electronically induced vibrational excitation, we focus on the electronic population and driving induced friction shown in Fig.~\ref{fig:Equilibrium_non-interacting}.
Note that for the model studied here, the population is related to the work rate by a factor of  $\dot \epsilon_d$ (here $\dot \epsilon_d = v$): 
\begin{eqnarray}
\dot W = \sum_i \dot R_i tr (\partial_i H  \rho ) =  \dot \epsilon_d tr (d^\dagger d \rho  ) =\dot \epsilon_d \langle d^\dagger d \rangle.
\end{eqnarray}
Consequently, the nonadiabatic work rate correction is related to the electronic friction by the following 
\begin{eqnarray}
\dot W^{(2)} =  \dot \epsilon_d^2 \gamma = \dot \epsilon_d^2  \delta \langle d^\dagger d \rangle, 
\end{eqnarray}
see also Eq.~\eqref{eq:gammadd}. 
In the following we focus on the electronic population, but keep in mind its simple relation with the work rate and its nonadiabatic correction.

When the electronic state energy is significantly below the chemical potentials, the electronic population is 1.
As the energy approaches and passes the value of the chemical potentials of the leads, the population drops down to 0. Accordingly, we observe a peak in the electronic friction centered at the position of the chemical potentials, i.e. $\epsilon_d(t)=\mu_L=\mu_R=0$.

The drop of the population and the friction peak are broadened by the temperature $T$ and by the environment-system coupling $\Gamma$. 
The coupling induced broadening is caused by co-tunneling processes, which are included in the HEOM results but disregarded in the analytical weak coupling result.
Since the friction obtained from QME or HEOM do barely show any differences even though $\frac{\Gamma}{k_\tB T}{=}1$, we conclude that 
the co-tunneling processes are not of great importance for the friction in the resonant-level-model case.
This is in contrast to what seems to happen in the presence of phonon coupling as discussed below.

The chosen finite driving velocities in the HEOM calculations lead to a slightly visible delay in the electronic population drop. For faster driving velocities, the delay becomes more pronounced and leads to a shift of the friction peak position in the direction of the driving. This friction peak shift is a nonadiabatic effect of higher order in the driving velocity.

\begin{figure}[!ht]
 \centering
\begin{minipage}{0.49\textwidth}
 \includegraphics{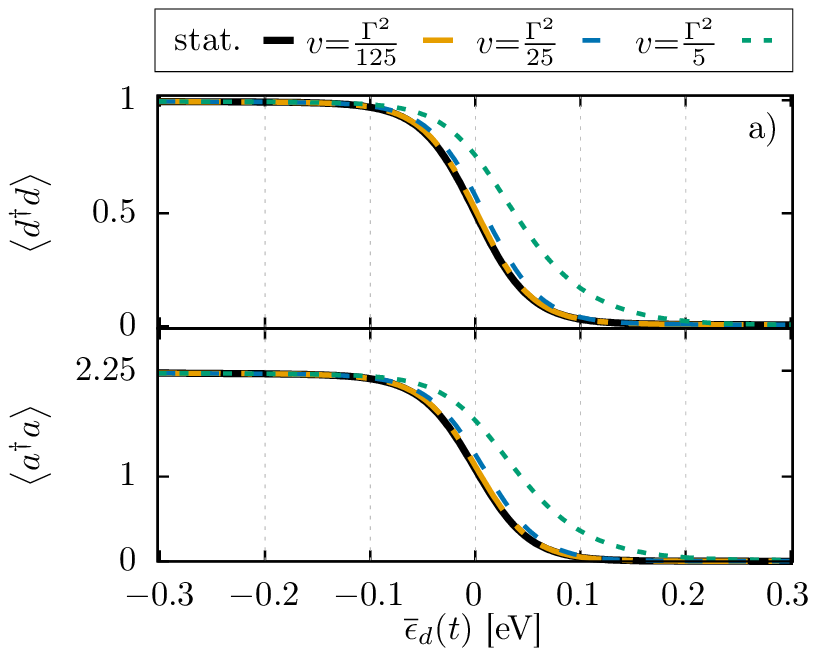}
\end{minipage}
\begin{minipage}{0.49\textwidth}
 \includegraphics{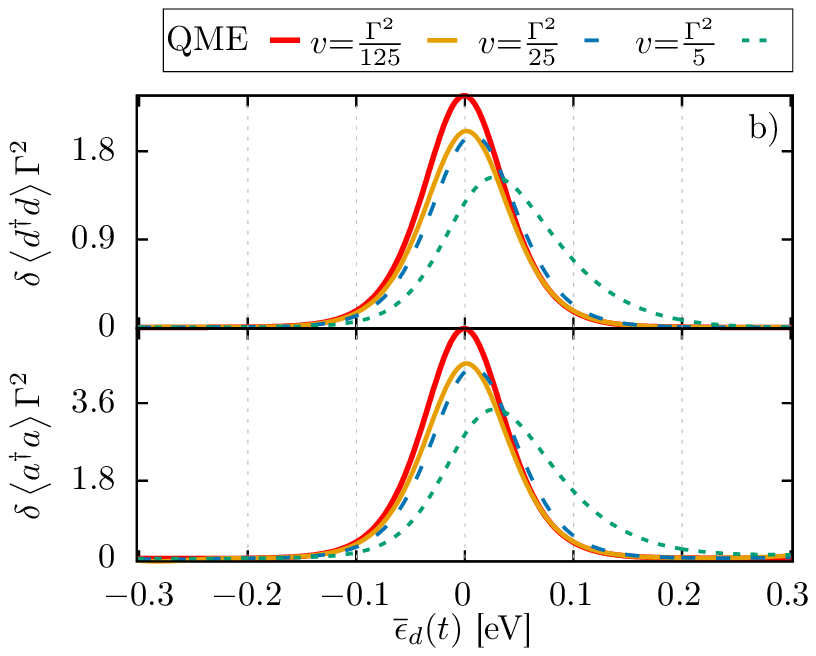}
 \end{minipage}
  \caption{ a) The system response (Electronic population $\av{d^\dagger d}$ and vibrational excitation $\av{a^\dagger a}$) and b) the corresponding nonadiabatic correction induced by drives with different velocities $v$ as a function of  the time-dependent energy $\overline{\epsilon}_d(t)=\overline{\epsilon}_0 + v t$. 
 The parameters are $\Phi=0$\,V, $\Omega=0.2$\,eV, $\frac{\lambda}{\Omega}=1.5$, $k_\tB T=0.025$\,eV, $\Gamma=0.025$\,eV, and $\Lambda=0$.}
\label{fig:Equilibrium_Interacting}
\end{figure}

\subsection{Zero bias voltage (equilibrium): Influence of electron-phonon coupling}
\label{sec:NumRes:sub:Eq:elph}
Next, we consider a system with strong electronic-vibrational interaction without applying a bias voltage, i.e. $\Phi=0$ and $\frac{\lambda}{\Omega}=1.5$.
In Fig.~\ref{fig:Equilibrium_Interacting} we depict the system response to linear driving of the electronic state energy and the corresponding nonadiabatic corrections   for different driving velocities. Here, we also plot the vibrational excitation, which is simply linked to the vibrational energy:
\begin{eqnarray}
E_v = \hbar\Omega \langle a^\dagger a \rangle  .
\end{eqnarray}
Therefore, the nonadiabatic contribution to the vibrational energy due to driving of $\epsilon_d$ is directly related to the nonadiabatic correction to the vibrational excitation:
\begin{eqnarray}
E_v^{(1)} = \dot \epsilon_d \hbar\Omega  \delta \langle a^\dagger a \rangle .
\end{eqnarray}

In the electronic population we observe the transition from the occupied state to the unoccupied state as the polaron shifted ground state energy of the system, $\overline{\epsilon}_d(t) = \epsilon_0-\frac{\lambda^2}{\Omega}+v t$, passes the chemical potentials. Apart from the clear visibility of the delay with faster driving velocity, the drop has a very similar form as in the non-interacting case (compare with Fig.~\ \ref{fig:Equilibrium_non-interacting}).

Due to the electronic-vibrational coupling, the change in the electronic population affects the vibrational equilibrium position and the vibrational  excitation.
As shown in Fig.\ \ref{fig:Equilibrium_Interacting}, the vibrational excitation drops in the same way as the electronic population from $\av{a^\dagger a}~=~\frac{\lambda^2}{\Omega^2}$ to $\av{a^\dagger a}~=~0$, which can be explained with in the small polaron picture.\cite{Mitra2004,Koch2005,Haupt2006,Leijnse2008,Haertle2011}  For the electronically unoccupied system the vibrational mode is not affected by the electronic-vibrational coupling. Hence, the unoccupied system has a vibrational equilibrium position $\av{a^\dagger + a }=0$, which allows for a vanishing vibrational excitation. On the other hand, for the electronically occupied system, the electron forms a polaron together with vibrational excitation which leads to a displaced equilibrium position $\av{a^\dagger + a }=\frac{2\lambda}{\Omega}$, where the minimal vibrational excitation is $\av{a^\dagger a}=\frac{\lambda^2}{\Omega^2}$.
Since the electronic state energy enforces the transition from electronically occupied to electronically unoccupied system, the vibrational excitation exhibits an according transition. At faster driving velocities, the delay in the dynamics of the vibrational excitation is similar to that in the electronic population.

In contrast to the non-interacting case of  Fig.~\ref{fig:Equilibrium_non-interacting}, in Fig.~\ref{fig:Equilibrium_Interacting}, the comparison of the electronic populations and the vibrational excitations for different driving velocities shows an obvious delay already for $v=\frac{\Gamma^2}{5}$. This reflects that the population dynamics is actually slower than indicated by $\Gamma$. The effective weakening of the electronic system-environment coupling has been discussed in detail by \textcite{Eidelstein2013} and is approximately described by $\Gamma_\text{eff}\approx \Gamma e^{-\frac{\lambda^2}{\Omega^2}}\approx 0.1 \Gamma$ for our parameters.

Concentrating on the HEOM results in Fig~\ref{fig:Equilibrium_Interacting}, the nonadiabatic corrections   for the electronic population and vibrational excitation exhibit a similar behaviour. They are both peaked around the chemical potential and shifted in the driving direction with increasing driving velocity.
We also note that the peak of the nonadiabatic correction is reduced at higher driving velocity (see the doted vs. the dashed and solid lines), however, this artifact is a result of the definition in Eq.~(\ref{eq:def_nonadiabatic_correction}) of the nonadiabatic correction which is divided by the driving velocity. This definition is inspired by a perturbation theory assuming a slow driving velocity and, as such, it fails to describe the nonadiabatic corrections   induced by fast driving velocities.  

Comparing the height of the peaks in Fig.~\ref{fig:Equilibrium_Interacting} and Fig.~\ref{fig:Equilibrium_non-interacting} we observe a factor of ${\sim}10$ increase in the latter. 
This enhancement of the nonadiabatic correction is induced by the effective reduction of the coupling strength of the system to the electronic environment. As a result, the time-scale for changes in the electronic population is prolonged.

In contrast to the resonant-level-model of Sec. \ref{sec:NumRes:sub:Eq}, we observe a significant difference in the nonadiabatic corrections   obtained by the QME and HEOM approach. The QME peaks are higher than the HEOM peaks.
Since such deviations are not visible in the non-interacting case and the QME approach does not capture co-tunneling processes, we conclude that 
the co-tunneling processes weaken the reduction of the effective electronic environmental coupling strengths.

\subsection{Non-vanishing bias voltage (nonequilibrium): Resonant-level-model }
\label{sec:NumRes:sub:NonEq}

\begin{figure}[!ht]
 \centering
\begin{minipage}{0.49\textwidth}
 \includegraphics{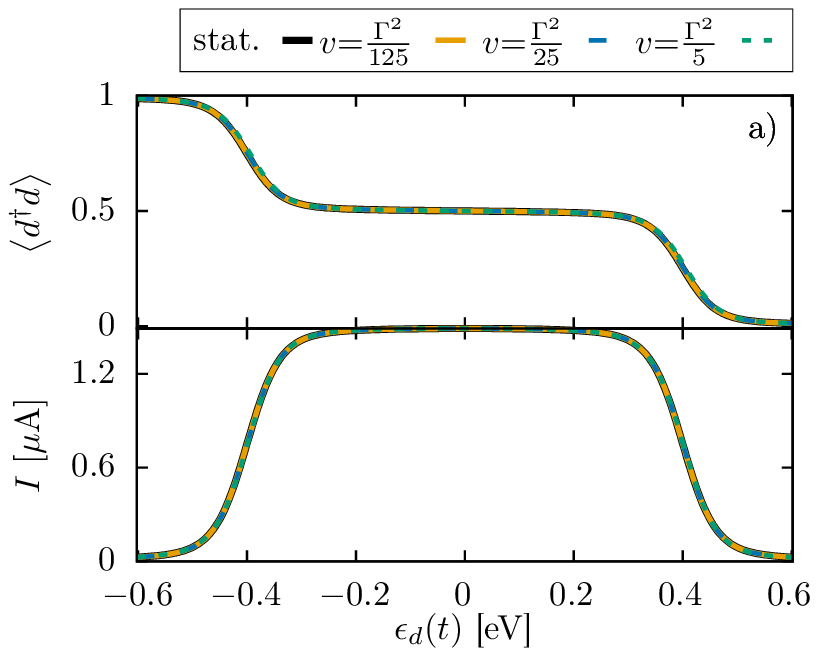}
\end{minipage}
\begin{minipage}{0.49\textwidth}
 \includegraphics{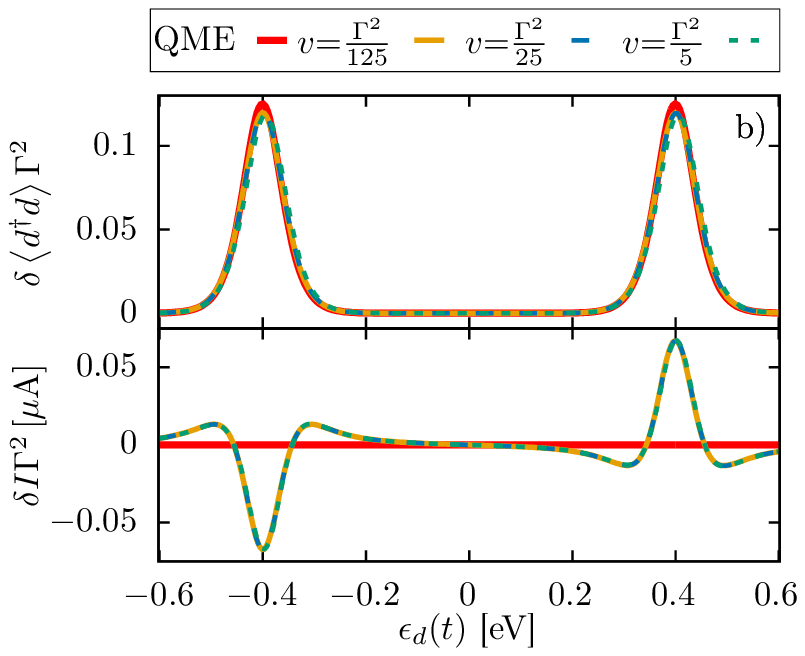}
\end{minipage}
 \caption{ a) The non-vanishing bias voltage system response (Electronic population $\av{d^\dagger d}$ and symmetrized charge-current $I=\frac{I_L-I_R}{2}$) and b) the corresponding nonadiabatic correction induced by drives with different velocities as a function of  the time-dependent energy ${\epsilon}_d(t)={\epsilon}_0 + v t$. 
 The parameters are $\Phi=0.8$\,V and $k_\tB T=\Gamma=0.025$\,eV.}
\label{fig:Bias0_8_non-interacting}
\end{figure}

In this section we investigate the electric friction and the nonadiabatic correction to the charge-current in the resonant-level-model where no electronic-vibrational coupling is present, however, the system is held out-of-equilibrium by applying a bias voltage symmetrically , i.e. $\Phi=2\mu_L=-2\mu_R$.
These quantities are plotted in Fig.~\ref{fig:Bias0_8_non-interacting}.
Since the two leads have different chemical potentials, the driven energy of the system is crossing their values one after the other, and as a result, the electronic population drops in two steps and the current is maximal between these two values,  $\mu_R\lesssim \epsilon_d(t)\lesssim \mu_L$. In this regime there are two competing processes of populating and depopulating the system by the two leads.

For the nonadiabatic correction of the electronic population, we find two positive peaks centered at the chemical potentials of the leads.
In this case, we observe a good agreement between the results obtained by the HEOM and QME. Yet, for the charge-current correction this is not the case, and we will explain the origin of this difference in more detail later.

The nonadiabatic correction to the charge-current in the HEOM approach is also peaked at the chemical potentials of the leads. However, the two main peaks have opposite signs.
Having a closer look at the peak structure, we notice that both main peaks are accompanied by smaller peaks in opposite directions. This peak structure looks similar to contributions to the differential conductance at the resonance due to co-tunneling.\cite{Koenig1997}
However, we find that these secondary peaks are also present in first tier truncated HEOM calculations, which include only sequential tunneling processes. Furthermore, they are still occurring for a weak system-environment coupling of $\frac{\Gamma}{k_B T}=\frac{1}{25}$ (see appendix \ref{sec:app:Details_sidepeaks}).
Therefore, 
our focus for explaining these smaller peaks is on sequential tunneling processes
. Without driving, the system obeys a time-translational and -reversal symmetry. Due to the time-translational symmetry, only resonant processes are determining the stationary state. By the linear energy shift both symmetries are broken and non-resonant processes are taking part in the dynamics.

\begin{figure}[!t]
 \centering
\begin{minipage}{0.49\textwidth}
 \includegraphics{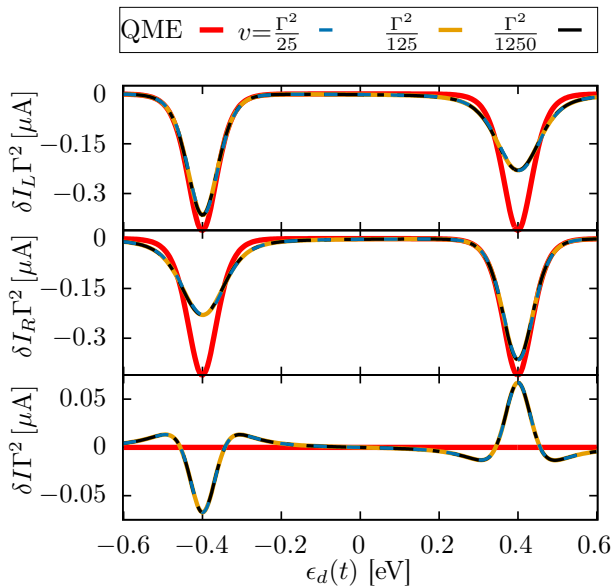}
\end{minipage}
 \caption{Nonadiabatic corrections   to different charge-currents for the non-vanashing bias voltage system.
 The parameters are $\Phi=0.8$\,V and $k_\tB T=\Gamma=0.025$\,eV.}
\label{fig:Bias0_8_currents_Friction_non-interacting}
\end{figure}

Non-resonant processes are especially important near the (de-)activation of resonant processes. In the following, we concentrate on the first chemical potential crossover (first peak of $\delta I$ from the left in Fig.~\ref{fig:Bias0_8_currents_Friction_non-interacting}) that represents an increase in the current. Before resonant processes become active, non-resonant processes contribute and increase the current. Thereby, these processes lead to the increase of the nonadiabatic correction before the resonant processes are dominating and we observe the main minimum.  After the resonant processes are activated, we again observe a current increase beyond the stationary value that is caused by the coherent superposition of non-resonant processes. For a more extreme situation of a sudden voltage change, these processes are known to induce the so-called "current ringing".\cite{Wingreen1993,Haertle2015}
The QME approach does not show any nonadiabatic corrections   to the charge-current. This suggest that our QME is not only excluding higher-order tunneling processes, but also ignores the contribution of non-resonant processes.

Since the symmetrized charge-current might not be accessible for experimentalists, we present the nonadiabatic correction to the current out of the individual leads $I_{L/R}$ as well as for the symmetrized charge-current in Fig.\ \ref{fig:Bias0_8_currents_Friction_non-interacting}.
We find that our QME approach predicts for both the left and right lead identical nonadiabatic corrections, resulting in a vanishing  correction to the symmetrized charge-current $I=\frac{I_L-I_R}{2}$. Furthermore, the nonadiabatic correction to the current of the left/right lead is proportional to the derivative of the stationary electronic population with respect to $\epsilon_d$. This is in agreement with the theoretical considerations  by \textcite{Splettstoesser2006} for an electron-electron interacting system.

In contrast, the HEOM approach reveals that the nonadiabatic correction peak in the current from a single electron reservoir is broader when its own chemical potential crosses the energy level of the system. 
This asymmetry in the peak heights and width in the nonadiabatic correction to the currents of the individual leads with respect to $\epsilon_d$ further supports the contribution of non-resonant processes to the nonadiabatic correction.
In appendix~\ref{sec:app:Details_sidepeaks} we present more details of the side peaks.

\begin{figure}[!b]
\centering
        \includegraphics{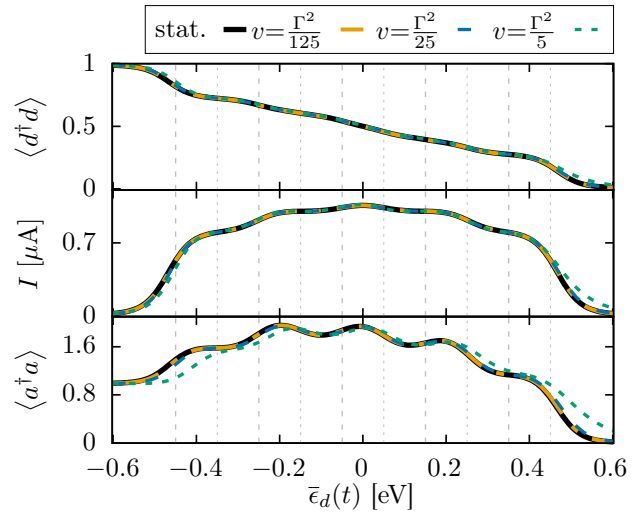}
        \caption{System response to linear drives in comparison to the stationary limit. Shown are the electronic population $\av{d^\dagger d}$, the vibrational excitation $\av{a^\dagger a}$, and the charge-current $I$ for different driving velocities $v$ as a function of the electronic state energy $\overline{\epsilon}_d(t)=\overline{\epsilon}_0 + v t$. By the dashed grey lines, we indicate the energetic position of the right chemical potential with an exchange of $n$ vibrational quanta $\mu_R +n \Omega$. The dotted grey lines indicate the the left chemical potential with an exchange of $m$ vibrational quanta $\mu_L - m \Omega$. Further parameters are $\Phi=0.9$\,V, $\Omega=0.2$\,eV, $\frac{\lambda}{\Omega}=1$, $k_\tB T=\Gamma=0.025$\,eV, and $\Lambda=0$.}
        \label{fig:Bias0_9_System_response}
\end{figure}

\subsection{Non-vanishing bias voltage (nonequilibrium): Influence of electron-phonon coupling }
\label{sec:Res:Bias_and_elph}
\begin{figure}[!ht]
\centering
        \includegraphics{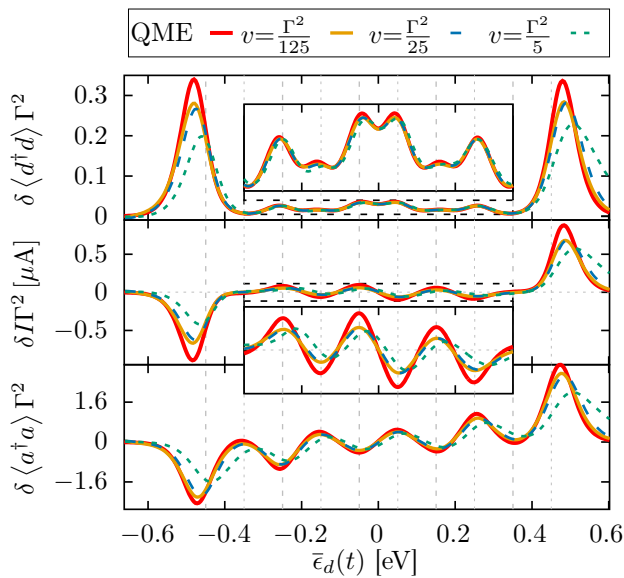}
        \caption{ Nonadiabatic correction  to the electronic population $\delta\av{d^\dagger d}$, the charge-current $\delta I$, and the vibrational excitation $\delta \av{a^\dagger a}$ for different driving velocities $v$ as a function of the electronic state energy $\overline{\epsilon}_d(t)=\overline{\epsilon}_0 + v t$. By the dashed grey lines, we indicate the energetic position of the right chemical potential with an exchange of $n$ vibrational quanta $\mu_R +n \Omega$. The dotted grey lines indicate the the left chemical potential with an exchange of $m$ vibrational quanta $\mu_L - m \Omega$. Further parameters are $\Phi=0.9$\,V, $\Omega=0.2$\,eV, $\frac{\lambda}{\Omega}=1$, $k_\tB T=\Gamma=0.025$\,eV, and $\Lambda=0$.}
        \label{fig:Bias0_9_Friction}
\end{figure}

Next, we add to the model of the previous section the electron-phonon coupling term. The results are illustrated in Fig.~\ref{fig:Bias0_9_System_response} and \ref{fig:Bias0_9_Friction}. 
We begin by focusing on the stationary state results with respect to the instantaneous Hamiltonian, the black continuous lines in Fig.~\ref{fig:Bias0_9_System_response}, that recovers the well understood Franck-Condon step structure with increasing the electronic state energy.\cite{Mitra2004,Galperinnano,Koch2005,Leijnse2008,Haertle2011}  

The electronic population decreases step wise from the completely occupied to the fully depleted electronic state for an energy range $\mu_R  \lesssim \overline{\epsilon}_d(t) \lesssim \mu_L $. While in this case the steps are only slightly visible, when considering the charge-current, the steps occur more distinct near the harmonic oscillator frequency shifted chemical potentials $\overline{\epsilon}_d(t)=\mu_{R/L}\pm n \Omega$. At these energies (dis-)charging processes with an exchange of $n$ vibrational quanta between the electronic reservoir ($R$) $L$ and the system become energetically (allowed) forbidden. 

Since the vibrational excitation is less frequently discussed in the literature, we discuss the onset and termination of different processes based on this observable in more detail here. 
Similar to the zero bias voltage case, the lowest vibrational excitation in the electronically occupied 
state 
is $\av{a^\dagger a}= \frac{\lambda^2}{\Omega^2}$. 

At energy values $\epsilon_d(t)=\mu_R+n\Omega$, discharging by the right lead with an additional excitation of the vibrational mode by $n$ quanta occurs, and thereby transport through the system becomes possible. 
Moreover, charging processes accompanied with an excitation of $m$ vibrational quanta become successively forbidden at $\overline{\epsilon}_d(t)=\mu_L-m\Omega$. 

For the chosen electronic-vibrational coupling $\frac{\lambda}{\Omega}=1$,
off-diagonal elements of the Franck-Condon matrix connected to processes with a large energy transfer into the vibrational mode are suppressed\cite{Koch2005,Leijnse2008,Haertle2011}. Hence, the according  steps are barely visible. 
For high energies of the electronic state, no charging process is energetically allowed and therefore the transport and the vibrational excitation vanish.
Overall, the alternating activation and deactivation of the different transport processes leads to the structured vibrational excitation as a function of the electronic state energy. 

In Fig.\ \ref{fig:Bias0_9_Friction}, we depict the nonadiabatic correction corresponding to the system response for different driving velocities. According to the step positions in the system response, we observe peaks in the nonadiabatic correction to the different physical quantities. 
In general, the nonadiabatic correction peaks illustrate the de-/activation of the transport processes described earlier. 
Here, we note that the nonadiabatic correction of the electronic population and the vibrational excitation are both explained by the delay in the underlying observables. In contrast, the nonadiabatic charge-current correction reflects a delayed dynamics for the peaks at $\overline{\epsilon}_d(t)\approx\mu_{R/L}$, but is ahead of the stationary limit at $\overline{\epsilon}_d(t) \in \{ \mu_R+n\Omega, \mu_L-m\Omega\}$ with $n,m \in \{1,2,3\}$.
Since a reduced vibrational excitation can enhance the charge-current\cite{Haupt2006}\footnote{This effect is also visible in Fig.~\ref{fig:Bias0_9_damping}.}, the delayed increase of the current-induced vibrational excitation causes the dynamic charge-current increase ahead of the stationary limit increase.
This is an example for the dynamical interplay 
of the charge-current and 
the present state of the vibrational mode.

Similar to the equilibrium case studied above, 
we observe a deviation between the peak heights obtained by HEOM and QME  which is caused by  co-tunneling processes in the weakening of the effective system-environment coupling.\footnote{See discussion in Sec.\ \ref{sec:NumRes:sub:Eq}.}

Moreover, focusing on  the peaks involving one or non vibrational quantum, we find clear deviations from the energetic position expected by the bare Polaron shift, i.e. $\overline{\epsilon}_d(t) \in \{ \mu_R+n\Omega, \mu_L-m\Omega\}$ with $n,m \in \{0,1\}$.
This renormalization effect is known for purely electronic interacting open quantum systems.\cite{Wunsch2005,Andergassen2010,Wenderoth2016} Here, we observe the system-environment coupling induced renormalization of an electronic-vibrational interacting open quantum system. We emphasize that we obtain this renormalization with in the numerically exact HEOM as well as the perturbative QME.

Next, we consider the higher-order nonadiabatic effects by comparing the nonadiabatic corrections  based on different driving velocity as shown in Fig. \ref{fig:Bias0_9_Friction}.
We find that the higher order nonadiabatic effects become apparent at a driving velocity of $v=\frac{\Gamma^2}{25}$ and are most pronounced at the energies with the largest peaks in the nonadiabatic correction.
For the fastest shown driving velocity of $v=\frac{\Gamma^2}{5}$, the higher order nonadiabatic effects in the current and the vibrational excitation are more pronounced at the smaller peaks than in the electronic population.

We recall that the effective coupling strengths between the system and the environment, which sets the time-scale for the electronic population dynamics, is weakened by the electronic-vibrational interaction.\cite{Eidelstein2013} 
For our parameters, the effective decay rate
is $\Gamma_\text{eff}= e^{-1}\Gamma$. Which means that the driving velocity of $v=\frac{\Gamma^2}{5}\approx1.48\Gamma^2_\text{eff}$ is already comparable to the time-scale of  the electronic population dynamics. From this perspective, a delay in the population dynamics is expected especially at energies $\overline{\epsilon}_d(t)$ with significant changes in the electronic population. However, the driving velocity is sufficiently slow that the population dynamics do not exhibit a significant delay for energies in the range $\mu_L-4\Omega < \overline{\epsilon}_d(t) < \mu_R + 4\Omega$, where the change in the electronic population with the electronic energy is smaller.

Including electronic-vibrational coupling, we no longer observe side peaks in the nonadiabatic correction to the charge-current. These side peaks are explained as contributions from coherent superposition of non-resonant processes. 
Therefore, the absence of the peaks means that the electronic-vibrational coupling induces a strong decoherence, which has already been reported in a different context.\cite{Haertle2013}

\subsection{Effect of vibrational relaxation}
\label{sec:NumRes:sub:EnvDamp}

\begin{figure}[!ht]
\centering
\begin{minipage}{0.49\textwidth}
\includegraphics{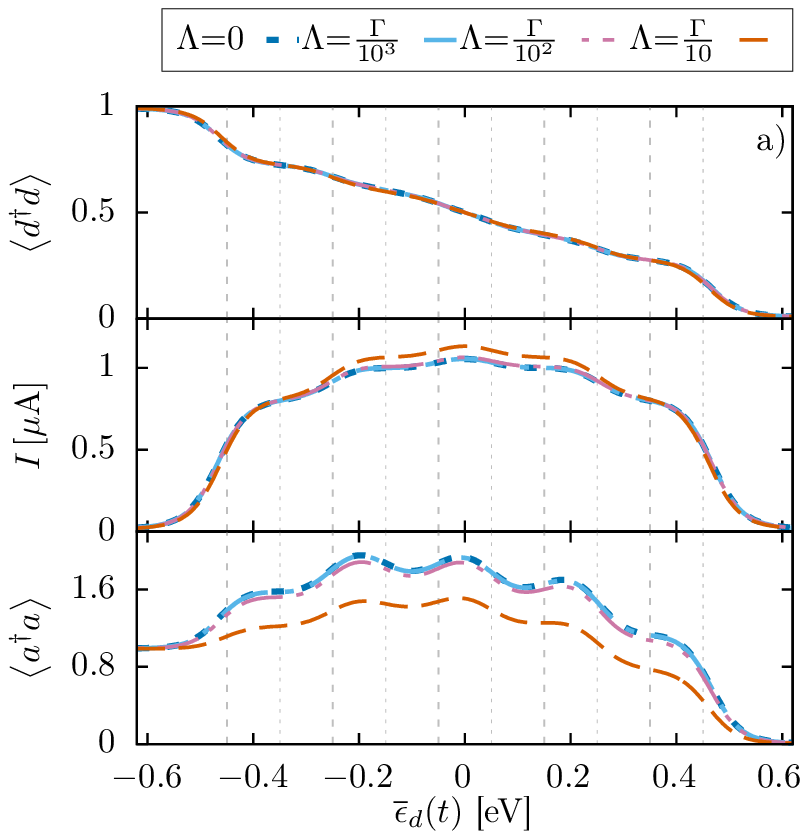}
\end{minipage}
\begin{minipage}{0.49\textwidth}
\includegraphics{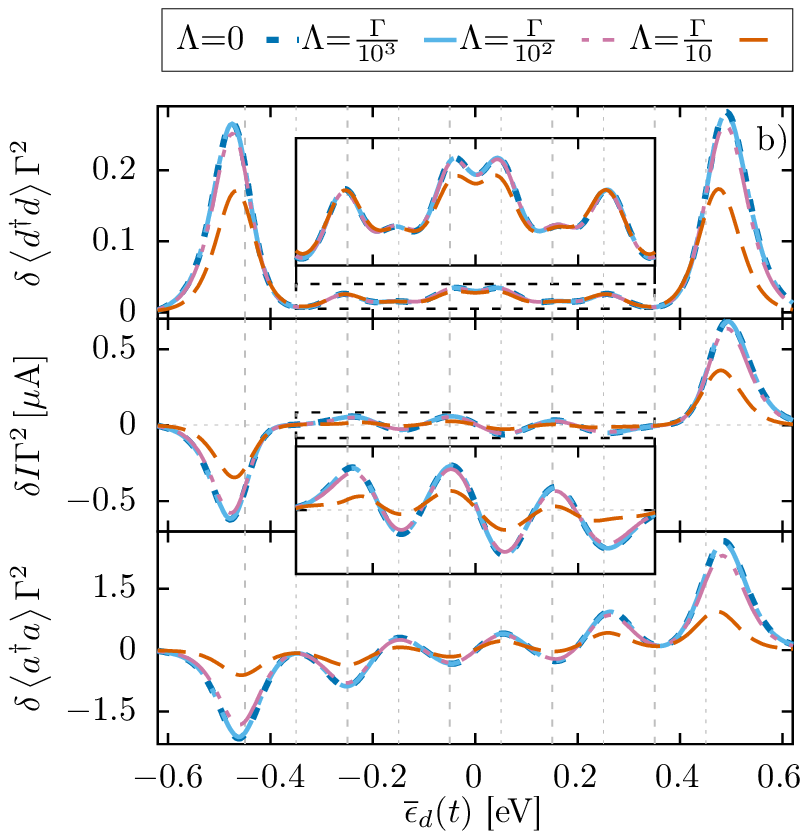}
\end{minipage}
\caption{ The effects of phonon damping on a) the electronic population $\av{d^\dagger d}$, charge-current $I$, and vibrational excitation $\av{a^\dagger a}$, and b) the corresponding nonadiabatic correction, as a function of  the time-dependent energy ${\epsilon}_d(t)={\epsilon}_0 + v t$ and for different phonon-bath coupling strength $\Lambda$.  Further parameters are $\Phi=0.9$\,V, $\Omega=0.2$\,eV, $\frac{\lambda}{\Omega}=1$, $k_\tB T=\Gamma=0.025$\,eV, and $v=\frac{\Gamma^2}{25}$.}
\label{fig:Bias0_9_damping}
\end{figure}

In the previous sections, we  disregarded the effect of a damping of the vibrational mode by the environment. In this section, we include this mechanism and investigate its effect on the nonadiabatic corrections  in Fig.~\ref{fig:Bias0_9_damping} for different values of the damping related coupling strengths $\Lambda$ and a relatively strong electronic-vibrational coupling $\frac{\lambda}{\Omega}=1$.

For $\Lambda \lesssim \frac{\Gamma}{10^2}$, we barely observe an effect of the damping on the stationary states (see  Fig.~\ref{fig:Bias0_9_damping}a) ) as well as the nonadiabatic correction (see Fig.~\ref{fig:Bias0_9_damping}b) ). Hence, the damping influence is weak in comparison to the current-induced vibrational excitation.
Its impact becomes clearly visible for the strongest damping strengths $\Lambda=\frac{\Gamma}{10}$. As expected in the limit $\Omega\ll k_\tB T$, the damping generally reduces the vibrational excitation. This effect is emerging particularly strong for energetic situations with a strong current-induced vibrational excitation.
Accordingly, we also observe a significant reduction of the nonadiabatic correction to the vibrational excitation.

As already reported in Refs. \onlinecite{Haupt2006,Baetge2021}, we notice an increase in the charge-current along with the decrease in the vibrational excitation. 
Furthermore, we observe a similar reduction of the nonadiabatic correction of the charge-current and of the vibrational excitation with increasing damping strengths, which again illustrates the dependence of the charge-current on the present state of the vibrational mode.

Since the highest peaks in the nonadiabatic correction of all observables, especially in the electronic population, are shifted towards $\mu_R$ and $\mu_l$, respectively, we conclude that 
the damping reduces the interaction induced renormalization
.

Similar to the significant effect of faster driving velocities to the highest nonadiabatic correction  peaks (see Fig.~\ref{fig:Bias0_9_Friction}), we find a clear reduction in the nonadiabatic correction in the highest nonadiabatic correction peaks of the electronic population. 
Here, the damping attenuates the reduction in coupling between the electronic system and the environment caused by the electronic-vibrational interaction. Consequently, this allows the population to adapt more quickly to the energetic situation and explains the decrease in the nonadiabatic correction.

\subsection{Fast driving}
\label{sec:NumRes:sub:FastDrive}

\begin{figure}[!b]
\centering
\begin{minipage}{0.495\textwidth}
 \includegraphics{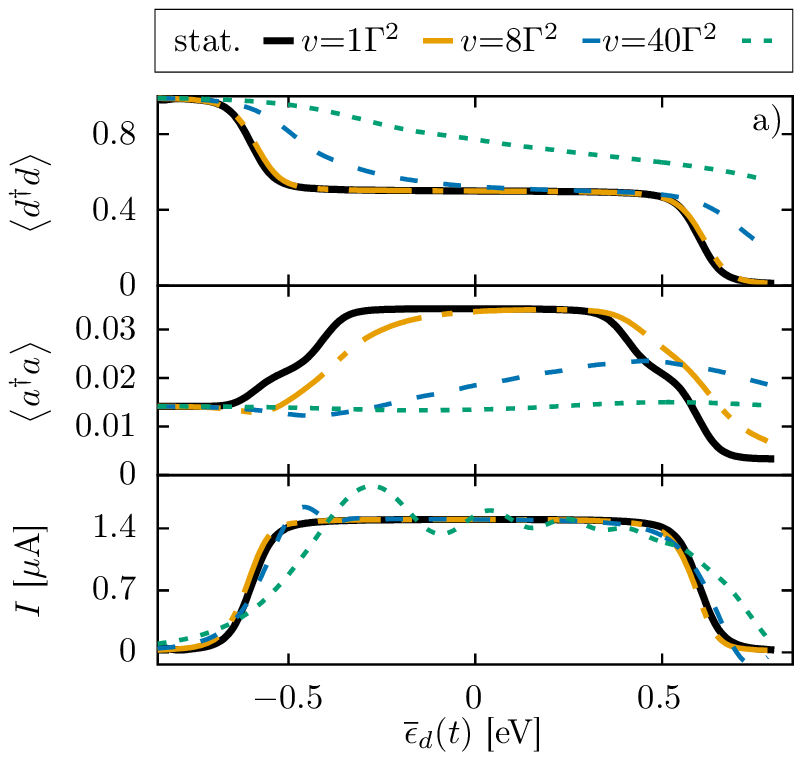}
\end{minipage}
\begin{minipage}{0.495\textwidth}
\includegraphics{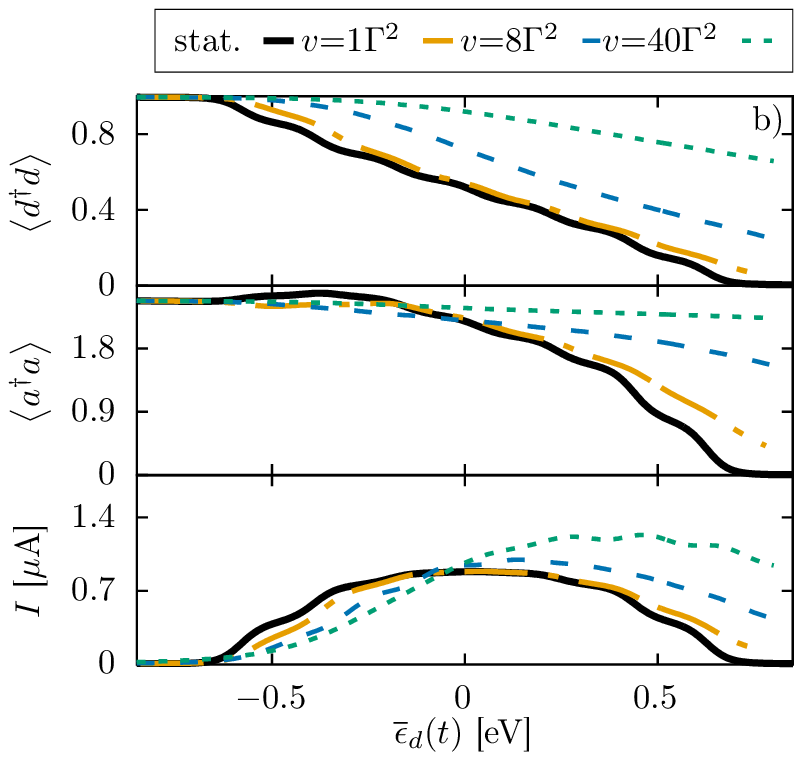}
\end{minipage}
\caption{ The system response (Electronic population $\av{d^\dagger d}$, charge-current $I$, and vibrational excitation $\av{a^\dagger a}$) as a function of  the time-dependent energy $\overline{\epsilon}_d(t)={\epsilon}_0 + v t$ for fast driving velocities and two different electronic-vibrational coupling $a) \frac{\lambda}{\Omega}=0.1$ and $ b) \frac{\lambda}{\Omega} =1.5$. 
Further parameters are $\Phi=1.2$\,V, $\Omega=0.2$\,eV, $k_\tB T=0.025$\,eV, $\Gamma=0.005$\,eV, and $\Lambda=0.025\,$eV.}
\label{fig:Bias1_2_fast_driving}
\end{figure}

In this section we present results for fast driving velocities.
As mentioned before, the nonadiabatic correction defined in Eq. \eqref{eq:def_nonadiabatic_correction} is no longer meaningful outside the linear response regime of slow driving.
Hence in Fig.~\ref{fig:Bias1_2_fast_driving}, we focus on the parameterized time-traces 
in comparison to their stationary state values for two different electronic-vibrational coupling strengths.
We begin our discussion with weak electronic-vibrational coupling $\frac{\lambda}{\Omega}=0.1$. For the driving speed $v=\Gamma^2$,  a significant delay is only visible in the time-trace of the vibrational excitation in comparison to its stationary state values. 
With increasing driving velocity, the delay increases visibly not only in the vibration excitation but also in the electronic population. In contrast, the delay occurring in the current is smaller. But additionally, the coherent superposition of the non-resonant processes leads to the so-called current ringing, which is known from instantaneous switches of the bias voltage.\cite{Wingreen1993}

Even though the electronic population decreases more uniform 
with increasing energy for a strong electronic-vibrational interaction $\frac{\lambda}{\Omega}=1.5$, the weakened effective system-environment coupling leads to a more obvious delay for $v=\Gamma^2$. 
At faster driving velocities, the delay in the electronic population and vibrational excitation increases in a manner qualitatively similar to the weak electronic-vibrational interaction.
In the current, we also observe an enhanced delay and an increase beyond the maximal steady state value. 
The latter is a clear fingerprint of non-resonant processes that coherently overlap despite the vibrationally induced decoherence, especially at fast driving velocities.


\section{Conclusions}
\label{sec:conclusion}

We have analyzed nonadiabatic corrections to thermodynamic properties for a non-equilibrium system under external modification. 
More specifically, we have investigated a system with and without electronic-vibrational interactions under a linear drive of the electronic state energy, utilizing the numerically exact HEOM as well as a perturbative quantum master equation approach. 

Without electronic-vibrational coupling, we found peaks in the friction and the nonadiabatic contribution to the charge-current when the system energy $\overline{\epsilon}_d$ crosses the chemical potentials of the electronic reservoirs. 
In accordance with the more complex onset and termination of transport processes induced by the electronic-vibrational interaction and understood in the Franck-Condon picture, we observed more complex responses in the nonadiabatic correction of the different quantities.
Furthermore, the results for different driving velocities, as well as different environmental vibrational damping, illustrated the
dynamical interplay of the electrons and the vibrational mode.

Surprisingly, the QME fails to reproduce the nonadiabatic current correction without electron-phonon coupling, while it qualitatively  recovers the nonadiabatic current correction in the adiabatic limit with electron-phonon coupling.
Moreover, the comparison of QME- and HEOM-based calculations reveals the contribution of co-tunneling processes to the electronic-vibrational, interaction-induced weakening of the system-environment coupling.

For transport scenarios with negligible electronic-vibrational coupling, our numerically exact approach reveals a significant effect of coherent non-resonant processes contributing to the nonadiabatic correction to the charge-current.
In contrast, we observe decoherence of the non-resonant processes in the adiabatic limit for systems with electronic-vibrational interaction.
Only for very fast driving velocities we do recover fingerprints of coherent non-resonant processes, including electronic-vibrational interaction related to the so-called "current ringing"\cite{Wingreen1993}.

\begin{acknowledgments}
It is a pleasure to acknowledge fruitful discussions with C. Kaspar, S. Rudge and S. Wenderoth.
WD acknowledges the startup funding from Westlake University. 
This work was supported by
the German Research Foundation (DFG) through a research grant and FOR 5099.
Furthermore, support by the state
of Baden-W\"urttemberg through bwHPC and the DFG
through Grant No. INST 40/575-1 FUGG (JUSTUS 2 cluster)
is gratefully acknowledged. This research was supported by the ISRAEL SCIENCE FOUNDATION (Grant No. 1364/21).

\end{acknowledgments}

\cleardoublepage
\appendix
\section{Details on the contribution of non-resonant processes to the nonadiabatic correction}
\label{sec:app:Details_sidepeaks}

In this appendix, we show more results supporting our discussion on the contribution of non-resonant transport processes to the nonadiabatic correction.

\begin{figure}[!t]
 \centering
 \includegraphics{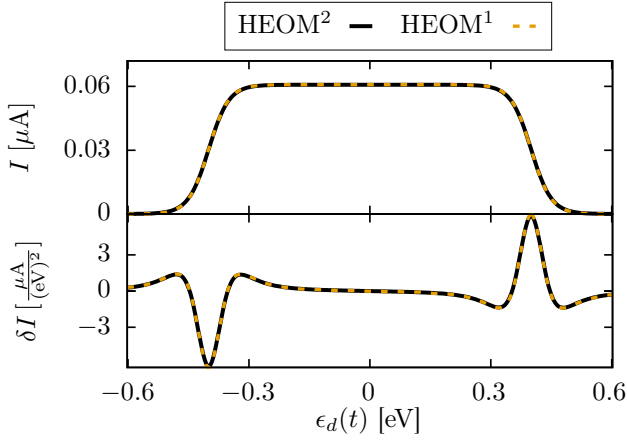}
 \caption{The charge-current $I$ through a non-interacting system and its nonadiabatic correction $\delta I$ as a function of the energy $\epsilon_d(t)$ obtained with in second/first tier truncated HEOM calculations denoted by HEOM$^{2/1}$. The parameters are $\Phi=0.8$\,V, $k_\tB T=0.025$\,eV and $\Gamma=0.001$\,eV.}
 \label{fig:app:First_order_check}
\end{figure}
In Fig.\ \ref{fig:app:First_order_check}, we show explicitly a comparison of HEOM results with a truncation in first and second tier. In first tier calculations only sequential tunneling processes are included and higher order processes like co-tunneling processes are excluded. We emphasize that second tier calculations are exact for non-interacting systems.\cite{Jin2008}
For the chosen weak coupling $\Gamma=0.04k_\tB T$, we do not observe a visible difference on the natural scale of the plot. Thereby, we validated our statement on the visibility of the side peaks in the nonadiabatic correction to the charge-current within first tier HEOM calculations.

\begin{figure}[!t]
 \centering
 \includegraphics{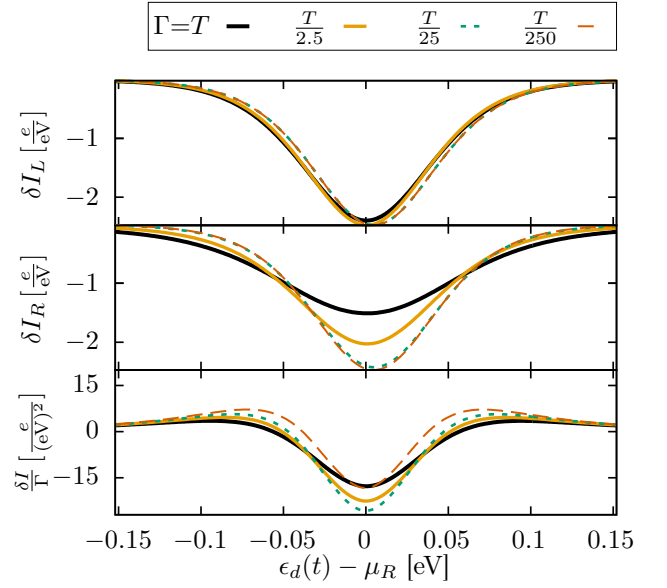}
 \caption{The nonadiabatic correction to different charge-currents  as a function of the energy $\epsilon_d(t)$ for different system-environment coupling strengths $\Gamma$. The parameters are $\Phi=0.8$\,V and $k_\tB T=0.025$\,eV. }
 \label{fig:app:Gamma_check}
\end{figure}
\begin{figure}[!b]
 \centering
 \includegraphics{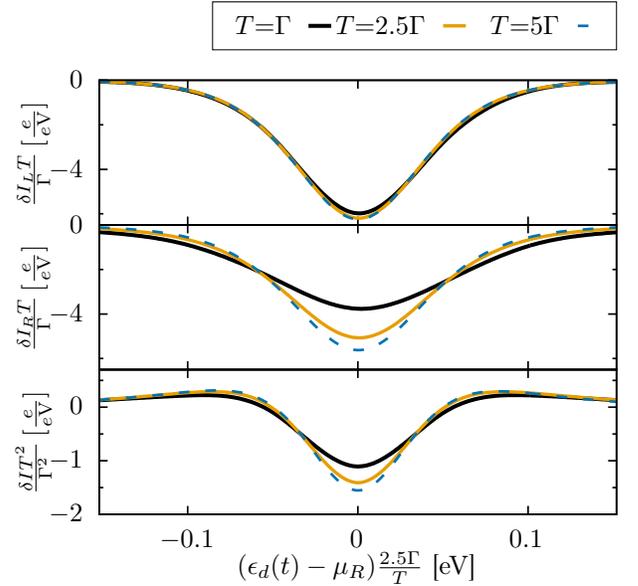}
 \caption{The nonadiabatic correction to different charge-currents  as a function of the energy $\epsilon_d(t)$ for different environmental temperatures $T$.  The parameters are $\Phi=0.8$\,V and $\Gamma=0.01$\,eV.}
  \label{fig:app:Temperature_check}
\end{figure}

Next, we demonstrate the dependence of the side peaks in the symmetrized charge-current on the system-environment coupling strength $\Gamma$ and the environmental temperature $T$ in Fig.\ \ref{fig:app:Gamma_check} resp. \ref{fig:app:Temperature_check}. Our results demonstrate, that the nonadiabatic correction to the currents of the individual electron reservoirs becomes more similar with decreasing $\Gamma$ and increasing temperature. Moreover, the nonadiabatic correction to the symmetrized charge-current vanishes roughly linearly with decreasing $\Gamma$ and quadratically with increasing $T$. Furthermore, the temperature leads to a rouhgly linear broadening of the peak.
Overall we emphasize, that the visibility of the non-resonant processes remains.
The slowest driving speed corresponds to $0.7596\frac{eV}{ns}$.

\begin{figure}[!ht]
\centering
 \begin{minipage}{0.49\textwidth}
 \includegraphics{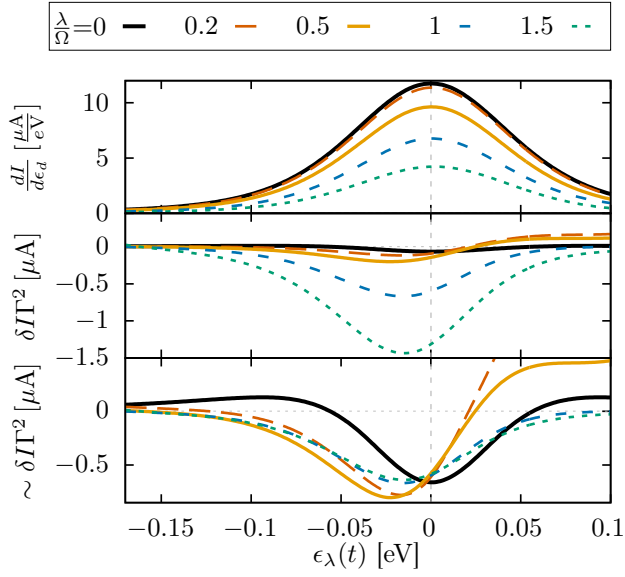}
 \end{minipage} 
 \caption{ The derivative of the stationary charge-current with respect to $\epsilon_d$, the nonadiabatic correction to the charge-current and rescaled nonadiabatic corrections   to the charge-current as a function of an aligned energy $\epsilon_\lambda (t) = \epsilon_d(t) +C_\lambda $. The $\lambda$-dependent energy shift $C_\lambda$ ensures that the first peak in the derivative of the stationary charge-current with respect to the energy is located at $\epsilon_\lambda(t)=0$.
 The parameters are  $\Phi=0.9$\,V, $\Omega=0.2$\,eV, $k_\tB T=\Gamma=0.025$\,eV, and $\Lambda=0$.}
  \label{fig:app:lambda_check}
\end{figure}
At last, we show the effect of the decoherence on the side-peaks for different electronic-vibrational interaction strengths $\lambda$ in Fig.\ \ref{fig:app:lambda_check}. Even for the weakest shown electronic-vibrational interaction $\frac{\lambda}{\Omega}=0.2$, we can not find a clear fingerprint of the coherent superposition of non-resonant processes in the nonadiabatic correction to the charge-current.
 Since the internal dynamics of the system becomes slower with decreasing $\lambda$, the HEOM calculations become numerically expensive.

\cleardoublepage

\bibliography{MyBib.bib}

\end{document}